\documentclass{JHEP3} 

\usepackage{epsfig,multicol,bbm}
\usepackage{cite}

\newcommand\fverb{\setbox\fverbbox=\hbox\bgroup\verb}
\newcommand\fverbdo{\egroup\medskip\noindent%
            \fbox{\unhbox\fverbbox}\ }
\newcommand\fverbit{\egroup\item[\fbox{\unhbox\fverbbox}]}
\newbox\fverbbox

\newcommand{\ppkkw}{$pp \to \bar{\chi} \chi +W^{\pm}+X$}

\title{ Dark matter pair associated with a $W$ boson production at the LHC in next-to-leading order QCD }

\author{Song Mao$^{(a)}$, Li Gang$^{(a)\ast}$, Ma Wen-Gan$^{(b)}$, Zhang Ren-You$^{(b)}$ and Guo Jian-You$^{(a)}$ \\
$^a$School of Physics and Material Science, Anhui University, Hefei, Anhui 230039, P.R.China \\
$^b$Department of Modern Physics, University of Science and
Technology of China, Hefei, Anhui 230026, P.R.China \\
E-mail: \email{songmao@mail.ustc.edu.cn}, \email{lig2008@mail.ustc.edu.cn}, \email{mawg@ustc.edu.cn},
\email{zhangry@ustc.edu.cn}, \email{jianyou@ahu.edu.cn}. }

\abstract{ We investigate the QCD next-to-leading order (NLO) corrections to the production of a pair of fermionic
dark matter particles associated with a W boson production through a mediator which couples to
standard model particles via either a vector or axial-vector coupling at the LHC.
We find that the QCD NLO corrections reduce the dependence of the total cross sections on the
factorization and renormalization scales, and the $K$-factors increase with the increment of the dark
matter mass. We also provide the LO and QCD NLO corrected distributions of the transverse momenta $p_T^\mu$ of final muon and transverse mass $M_T$.
We find that the LO cross section is significantly changed by the QCD NLO corrections. }


\vfill \eject

\begin{document}

\par
\section{Introduction}
\par
Observational evidence has confirmed that there exist some kinds of
cold non-baryonic dark matter (DM) in our universe, which is the dominant
component of cosmical matter \cite{Bertone04}. Due to the
characteristics of non-baryonic, the dark matter cannot be composed
of anyone of known substances in our earth and galaxies.
Astrophysical observations also can not tell us about the property
of the dark matter particle or whether it interacts with the
Standard Model (SM) particles beyond gravitation. Revealing the
distribution and nature of dark matter is one of the most
interesting current challenges in the fields of both cosmology and
particle physics.

\par
Among all the dark matter candidates, weakly interacting massive
particle (WIMP) is one of the most compelling versions. Primarily
this is due to that it offers a possibility to understand the
relic abundance of dark matter as a natural consequence of the
thermal history of the universe\cite{Feng:2008}. Some extensions of the SM, such as Supersymmetry
\cite{Martin:1997ns,Drees:2004jm}, Universal Extra Dimensions
\cite{Appelquist:2000nn} or Little Higgs Models
\cite{ArkaniHamed:2001nc}, naturally lead to good candidates for
WIMPs and the cosmological requirements for the WIMP abundance in
the universe. In these theoretical frameworks, the WIMP candidates
are often theoretically well motivated and compelling. However, all
of these theories still lack experimental support, and we can not
judge which theory is proper for dark matter.
Additionally, the first observations of dark matter may come from
direct- or indirect-detection experiments, which may not provide
information about the general properties of the dark matter particle
without offering a way to distinguish between underlying theories.
Thus, model independent studies of dark matter phenomenology using
effective field theory is particularly important.

\par
Recently, the observed results favor a light DM with a mass around
$10~{\rm GeV}$ in various experiments, such as DAMA, CoGeNT and
XENON10/100 \cite{Bernabei:2010mq,Aalseth:2010vx,Aprile:2010um,Sorensen:2010hq}.
However, it is difficult to probe in the low-mass region or to
constrain the parameter space with direct detection experiments,
since the typical energy transfer in the scattering of such
particles is small compared to the experimental energy thresholds,
and the interpretation of experimental results are affected by
astrophysical uncertainties. As colliders are most effective when producing highly boosted light WIMPs,
one can directly probe the interactions
between DM particles and individual SM particles at
the LHC. In the case of a WIMP, stability on the order of the
lifetime of the universe implies that pair production must highly
dominate over single production, and precludes the WIMP from
decaying within the detector volume. Therefore, WIMPs appear as
missing energy, and can potentially be observed by searching for
visible particles recoiling against dark matter particles
\cite{Birkedal:2004xn,Beltran:2008xg,Cao:2009uw,Beltran:2010ww,Shepherd:2009sa}

\par
Searches for dark matter in missing momentum channels can be
classified based on the visible particles against which the
invisible particles recoil. Existing experimental studies have
considered the cases in which the visible radiation is a jet of
hadrons (initiated by a quark or
gluon)~\cite{cdfjmet,ATLAS:2012ky,Chatrchyan:2012me}, a
photon~\cite{Aad:2012fw,Chatrchyan:2012tea}, or a $W/Z$ boson
decaying into leptons or hadronic jets
~\cite{TheATLAScollaboration:2013fia,CMS:2013iea,Aad:2013oja}.
Because the LHC is a proton-proton collider, the QCD correction
should be considered for any process if people want to make a
reliable prediction. More recently, The production of DM pairs plus
a jet or photon have been calculated to QCD next-to-leading order (NLO)
\cite{Haisch:2012kf, Fox:2012ru,Huang:2012hs,Wang:2011sx,Haisch:2013ata}. However, the
production of DM pairs in association with a mono-$W/Z$ has only
been considered to LO
~\cite{Bai:2012xg,Carpenter:2012rg,Bell:2012rg}. In this work, using
model independent method we investigate the possibility of
discovering the DM production in associated with a $W$ boson
induced by a dimension six effective operator in the next-to-leading
order (NLO) QCD.

\par
The paper is organized as follows. In section 2, we briefly describe
the related effective field theory and present the details of
calculation strategy. In section 3, we give the numerical results and discussion for the
process \ppkkw. Finally, a short summary
is given in section 4.

\vskip 5mm
\section{Calculation descriptions}
\par
We assume that the dark matter candidate is the only new particle
which is singlet under the SM local symmetries, and all SM particles
are singlets under the dark-sector symmetries. The interaction
between the SM and DM sectors is presumably effected by the exchange
of some heavy mediators whose nature we do not need to specify, but
only assume that they are much heavier than the typical scales. The
effective field theories for dark matter interacting primarily with
SM quarks have been considered in
Refs.~\cite{Beltran:2008xg,Shepherd:2009sa,Cao:2009uw,Beltran:2010ww,
Goodman:2010yf,Bai:2010hh,Goodman:2010ku,Rajaraman:2011wf,Fox:2011pm,Cheung:2012gi}.
The most prominent coupling characters are:
\begin{eqnarray}
{\rm Spin-independent~vector~coupling:(V)}&~\frac{1}{\Lambda_{\rm D5}^2}
\left(\bar{\chi} \gamma^\mu \chi\right) \left(\bar{q} \gamma_\mu q\right) \\
{\rm
Spin-dependent~axial-vector~coupling:(AV)}&~\frac{1}{\Lambda_{\rm
D8}^2} \left(\bar{\chi} \gamma^\mu \gamma^5 \chi\right)
\left(\bar{q} \gamma_\mu \gamma^5 q\right), \label{eq:EFTq}
\end{eqnarray}
where $\chi$ is the dark matter particle, which we assume to be a Dirac fermion,
$q$ is a SM quark, and the characterizing parameters of the model are the scales of the effective
interactions $\Lambda_{i}=\frac{M_{messenger}}{\sqrt{g_\chi g_q}}$ between the two sectors.
We will typically consider only one interaction type to
dominate at a time, and will thus keep one $\Lambda$ being finite while the rest
is set to be infinity and decoupled.
\begin{figure}[!htb]
\begin{center}
\begin{tabular}{cc}
{\includegraphics[width=10cm]{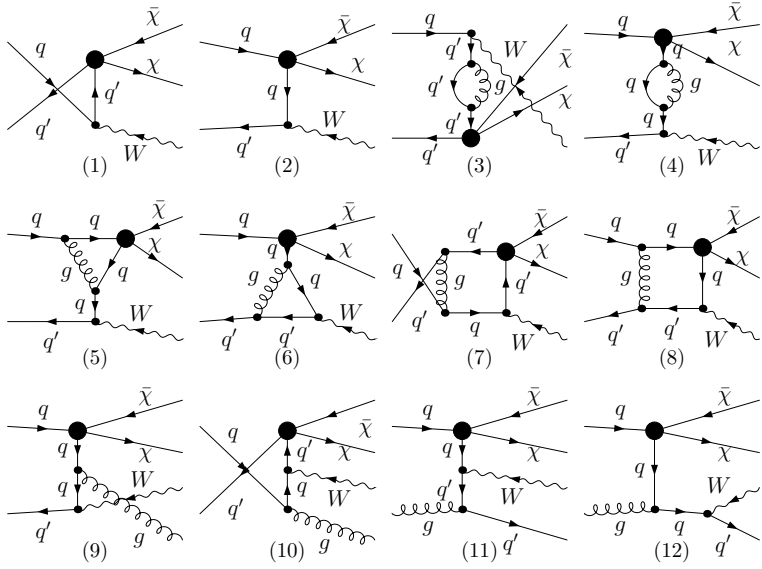}}
\end{tabular}
\end{center}
\vspace*{-0.5cm}\caption{\label{f1} Some representative Feynman diagrams for the process \ppkkw at the LO and
QCD NLO.}
\end{figure}

\par
There are two different topological
diagrams for process \ppkkw ~at the LO, in which dark matter pair of
WIMPs ($\bar{\chi}\chi$) produced via an unknown intermediate state,
with initial state radiation of a $W$ boson. Due to the
$CP$-conservation, the cross section for the $qq' \to \bar{\chi} \chi
+W^{-} ~(qq' = \bar{u}d,\bar{u}s,\bar{c}d,\bar{c}s)$
subprocess should be the same as that the corresponding charge
conjugate subprocess $qq' \to \bar{\chi} \chi +W^{+} ~(qq' =
u\bar{d},u\bar{s},c\bar{d},c\bar{s})$ at the parton level. By taking
the Cabibbo-Kobayashi-Maskawa (CKM) matrix elements $V_{td} = V_{ts}
= V_{ub} = V_{cb} = 0$, the LO contribution to the cross section for
the parent process \ppkkw ~comes from the following subprocesses. We
denote the subprocesses as:
\begin{equation}
\label{process1} q(p_1)+ q'(p_2) \to \bar{\chi}(p_3)+ \chi(p_4)+ W^{+}(p_5),~~~~(qq' = u\bar{d},u\bar{s},c\bar{d},c\bar{s})
\end{equation}
\begin{equation}
\label{process2} q(p_1)+ q'(p_2) \to \bar{\chi}(p_3)+ \chi(p_4)+ W^{-}(p_5),~~~~(qq' = \bar{u}d,\bar{u}s,\bar{c}d,\bar{c}s)
\end{equation}
where $p_{1}$, $p_{2}$ and $p_{3}$, $p_{4}$, $p_{5}$ represent the
four-momenta of the incoming partons and the outgoing dark matter particle $\chi$,
and $W^{\pm}$ boson, respectively.

\par
In the QCD NLO calculations, the parent process \ppkkw ~involves
four contribution components: (1) The QCD one-loop virtual
corrections to the partonic process $qq' \to \bar{\chi} \chi+
W^{\pm}$; (2) The real gluon emission partonic process $qq' \to
\bar{\chi} \chi + W^{\pm} + g$; (3) The real light-(anti)quark
emission partonic process $qg \to \bar{\chi} \chi + W^{\pm} + q'$;
(4) The corresponding contributions of the parton distribution
functions (PDF) counterterms.

\par
Some representative QCD NLO Feynman diagrams for the process $pp \to
\bar{\chi} \chi W^{\pm} +X~$ are shown in Fig.\ref{f1}. In virtual
correction calculations, we adopt the definitions of scalar and
tensor one-loop integral functions in Refs.
\cite{Passarino,denner2}. Using the Passarino-Veltman (PV) method
\cite{Passarino,denner3}, the tensor integrals are expressed as a
linear combination of tensor structures and coefficients, where the
tensor structures depend on the external momenta and the metric
tensors, and the coefficients depend on scalar integrals, kinematics
invariants and the dimension of the integral. After the tensor
integral reduction is performed, the fundamental building blocks are
one-loop scalar integrals. we adopt the dimensional regularization
(DR) method in $D = 4 - 2\epsilon$ dimensions to isolate the
ultraviolet (UV), soft infrared (IR), and collinear IR divergences.
The one-loop scalar integrals arise from IR divergent
box diagrams with several external and internal massless particles is most difficult.
We adopt the expressions in Ref.\cite{IRDV} to deal with the IR
divergences in Feynman integral functions, and apply the expressions
in Refs.\cite{OneTwoThree,Four,Five} to implement the numerical
evaluations for the IR safe parts of N-point integrals. The UV
divergences are removed by the counter terms fixed by the
on-mass-shell renormalization condition. Nevertheless, it still
contains soft/collinear IR singularities. The infrared (IR)
singularities from the one-loop integrals need to be cancelled by
adding the contributions of the real gluon emissions.

\par
The real gluon emission subprocess $qq' \to \bar{\chi} \chi W^{\pm} g$
contains both soft and collinear IR singularities which can be
conveniently isolated by adopting the two cutoff phase space slicing
(TCPSS) method \cite{TCPSS}. The soft divergences from real gluon emission will
be canceled by similar singularities from the
contribution of the one-loop diagrams. The collinear
divergences from real gluon emission corrections, part of it can be eliminated by
collinear singularities in virtual corrections, and the remaining
collinear divergences in real gluon corrections can be absorbed into
the PDFs. Then the UV and IR singularities are exactly vanished after combining the
renormalized virtual corrections with the contributions of the real
gluon emission processes and the PDF counterterms together.
These cancelations can be verified numerically
in our numerical calculations.

\par
The real light-quark emission subprocesses have the same order
contribution with previous real gluon emission subprocess.
These subprocesses contain only the initial
state collinear singularities. Using the TCPSS method, the phase
space is divided collinear region and non-collinear region. The
cross section in the non-collinear region is finite and can be
evaluated in four dimensions using the general Monte Carlo method.
While the collinear singularity in collinear region can be absorbed
into the redefinition of the PDFs at the NLO. We have implemented
all above calculations in the way as presented in our previous works
\cite{Mao:2013dxa,oai:arXiv.org:0903.2885}.

\vskip 10mm
\section{Numerical results and discussion}
\par
In this section we present the numerical results and discussions for
the $pp \to \bar{\chi}\chi+W^{\pm}+X~$ process at both the LO and
the QCD NLO. We take CTEQ6L1 PDFs with an one-loop running
$\alpha_s$ in the LO calculation and CTEQ6M PDFs with a two-loop
$\alpha_s$ in the NLO calculation \cite{CTEQ6}, and the
corresponding fitted values $\alpha_s(M_Z) = 0.130$ and
$\alpha_s(M_Z) = 0.118$ are used for the LO and NLO calculations,
respectively. For simplicity we define the factorization and
the renormalization scale being equal, and take $\mu\equiv\mu_f =
\mu_r = (2m_\chi+ m_W)/2$ by default unless stated otherwise. We
adopt all the quark masses $m_u=m_d=m_c=m_s=0$ and employ the
following numerical values for the relevant input parameters
\cite{pdg},
\begin{equation}
\begin{array}{lll}  \label{input1}
\alpha(m_Z)^{-1}=127.918, &m_W=80.398~{\rm GeV},    &m_Z=91.1876~{\rm GeV}. \\
\end{array}
\end{equation}
The CKM matrix elements are fixed as
\begin{equation}
\begin{array}{lll}  \label{input2}
V_{CKM}=\left(                 
\begin{array}{ccc}   
 V_{ud} & V_{us} & V_{ub}\\
 V_{cd} & V_{cs} & V_{cb}\\
 V_{td} & V_{ts} & V_{tb}\\
\end{array}\right)=
\left(                 
\begin{array}{ccc}   
 0.97418 & 0.22577 & 0\\
 -0.22577 & 0.97418 & 0\\
 0 & 0 & 1\\
\end{array}\right).
\end{array}
\end{equation}

\par
In order to verify the correctness of our tree-level calculation,
we compare our numerical results with those in Ref.\cite{CMS:2013iea} in the same input parameters.
We find that they are consistent within the allowable error range.
In the calculation of real corrections, we adopt the two-cutoff
phase space slicing method \cite{TCPSS}. The two phase space cutoffs
$\delta_s$ and $\delta_c$ are chosen as $\delta_s=10^{-3}$ and
$\delta_c=\delta_s/50$ as default choice. In checking the
independence of the final results on two cutoffs $\delta_s$ and
$\delta_c$, we find the invariance with the $\delta_s$ running from
$10^{-2}$ down to $10^{-4}$ within the error control.
\begin{figure}
\centering
\includegraphics[width=0.8\textwidth]{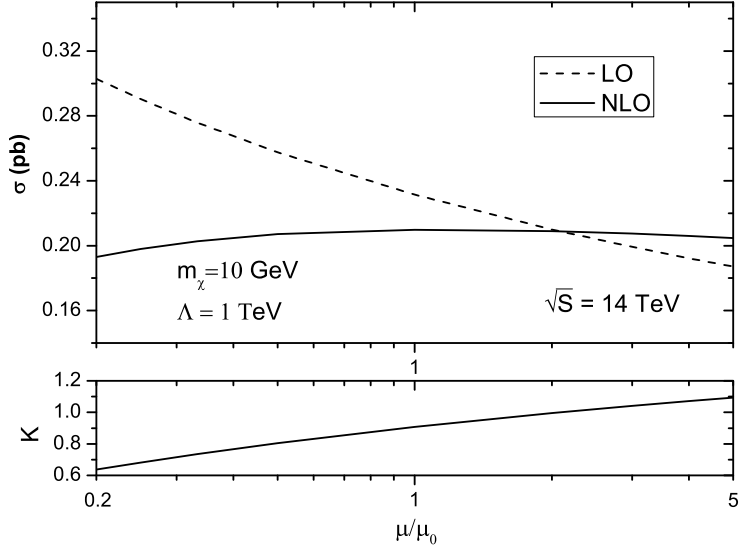}
\vspace{0cm}
\caption{\label{fig2} The dependence of the LO, NLO QCD corrected
total cross sections and the corresponding $K$-factors
($K(\mu)\equiv \sigma_{NLO}(\mu)/\sigma_{LO}(\mu)$) on the
factorization/renormalization scale ($\mu/\mu_0$) for the process
$pp \to \bar{\chi} \chi +W^{\pm}+X$ at the ${\sqrt{s}=14}$ TeV LHC
by taking $m_\chi=10~GeV$ and $\Lambda=1~TeV$. Here we assume
$\mu=\mu_r=\mu_f$ and define $\mu_0 = (2 m_\chi + m_W)/2$.  }
\end{figure}

\par
In Fig.\ref{fig2}, we illustrate the renormalization/factorization
scale dependence of the LO, NLO QCD corrected total
cross sections and the corresponding $K$-factors ($K(\mu)\equiv
\sigma_{NLO}(\mu)/\sigma_{LO}(\mu)$) for the process $pp \to
\bar{\chi}\chi+W^{\pm}+X$. We assume $\mu=\mu_r=\mu_f$ and define $\mu_0 = (2 m_\chi + m_W)/2$,
where the mass of DM and coefficients $\Lambda$ are taken as $m_\chi=10~GeV$ and $\Lambda=1~TeV$.
Since the difference of theory predicts induced by the vector
(D5) operator and axial-vector (D8) is very small, the lines
for D5 and D8 in Fig.\ref{fig2} look essentially the same that
it is difficult to distinguish between them.
We can see that the dependence of the NLO cross section on the
factorization/renormalization scale are
significantly reduced comparing with the LO cross section. This
makes the theoretical predictions much more reliable. When the
renormalization/factorization scale varies from $0.2 \mu_0$ to
$5 \mu_0$, the corresponding $K$-factor increases from $0.64$ to
$1.09$ for both the vector (D5) and axial-vector (D8) operator.

\begin{figure}
\centering
\includegraphics[width=0.7\textwidth]{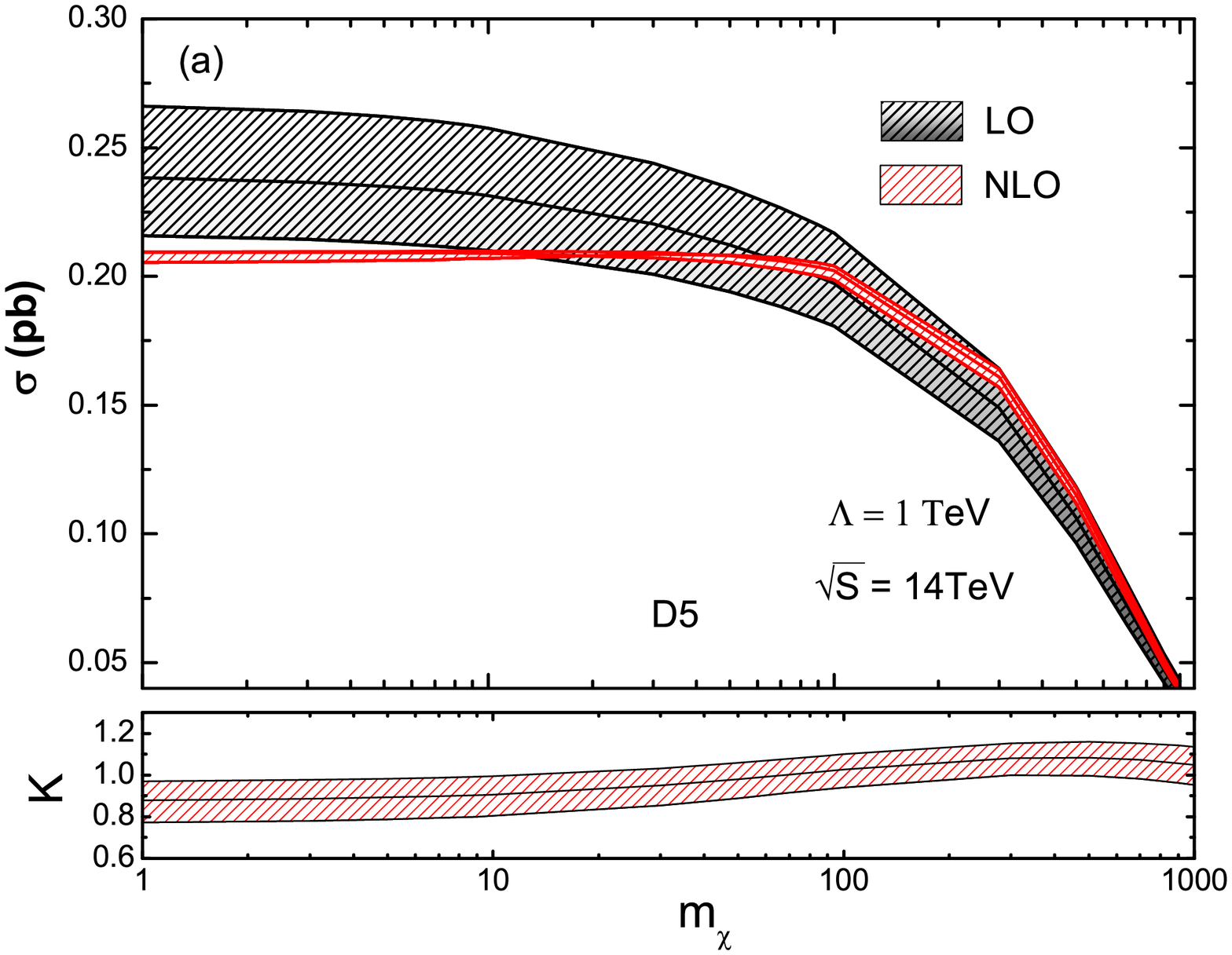}
\includegraphics[width=0.7\textwidth]{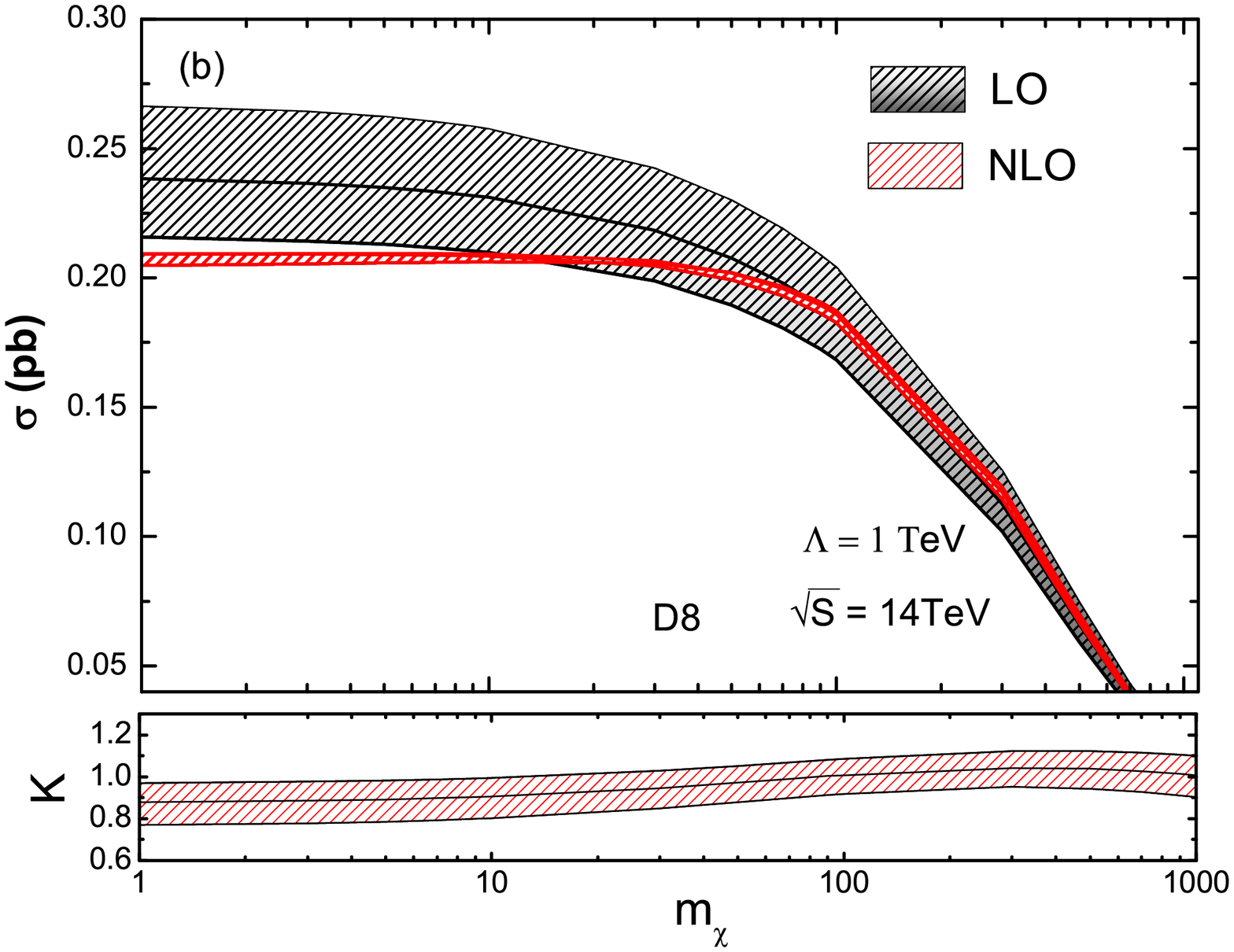}
\caption{\label{fig3} The LO, QCD NLO corrected integrated cross sections and the
corresponding $K$-factors as the functions of the DM mass for DM production in
association with mono-$W$ boson at the $14~TeV$ LHC with $\Lambda=1~TeV$ with the default scale $\mu=\mu_0$,
the shaded band represents the deviation from this scale when the scales are varied by a factor of two in each direction.
(a) for the vector (D5) ~~(b) for the axial-vector (D8) }
\end{figure}

\par
In Fig.\ref{fig3}, we show the DM mass dependence of the LO and QCD
NLO corrected cross sections for the process $pp \to \bar{\chi} \chi
+W^{\pm}+X$ with $\Lambda=1~TeV$ at the $14~TeV$ LHC induced by the
vector (D5) and axial-vector (D8) operator. The dependence on the
renormalization and factorization scales, taken to be equal
($\mu=\mu_r=\mu_f$), is illustrated by the shaded band linking the
predictions obtained at $\mu=2 \mu_0$ and $\mu=1/2 \mu_0$, whilst
the central scale choice $\mu=\mu_0$ is illustrated by the curve
inside the shaded band. When the DM mass varies from $1~GeV$ to
$1000~GeV$, the QCD NLO corrections modify the LO cross sections
obviously. As we expected, the overall scale dependence up to NLO is
smaller than the LO prediction. In the DM mass range from $1$ to
$90~GeV$, we see that the QCD NLO corrections generally reduce the
LO cross sections particularly significant in the lower DM mass
range, and the $K$-factors increase from $0.88$ to $1.0$ for both
the vector (D5) and axial-vector (D8) operator. Thus, it is
necessary to consider the QCD NLO corrections for the process of DM
production at hadron colliders. The contributions of the vector (D5)
and axial-vector (D8) operator can be distinguished until the DM
mass being more than $100~GeV$. The similar behavior is demonstrated
in the monojet production at the LHC \cite{Fox:2012ru}. The vector
and axial-vector DM operators show similar behaviour in terms of
K-factors and scale dependence. This is because that in the massless
limit the only terms which are sensitive to the axial nature of the
coupling are the four-quark amplitudes, which are a small part of
the total NLO cross section. As the the DM mass grows, so does the
difference between the operators, with the axial operator being
smaller than the vector over the entire mass range.

\par
\begin{figure}
\centering
\includegraphics[width=7.5cm,height=6cm]{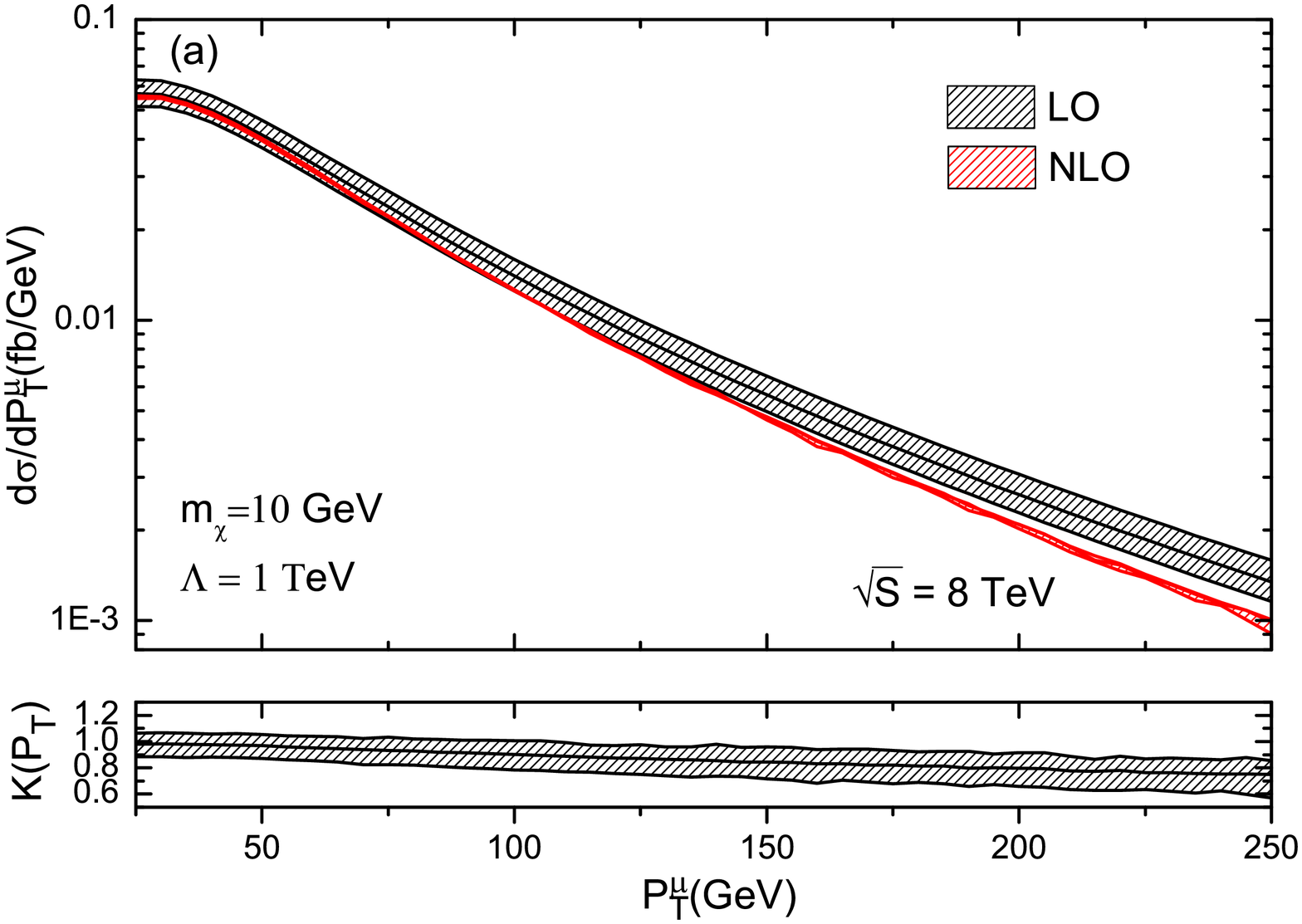}
\includegraphics[width=7.5cm,height=6cm]{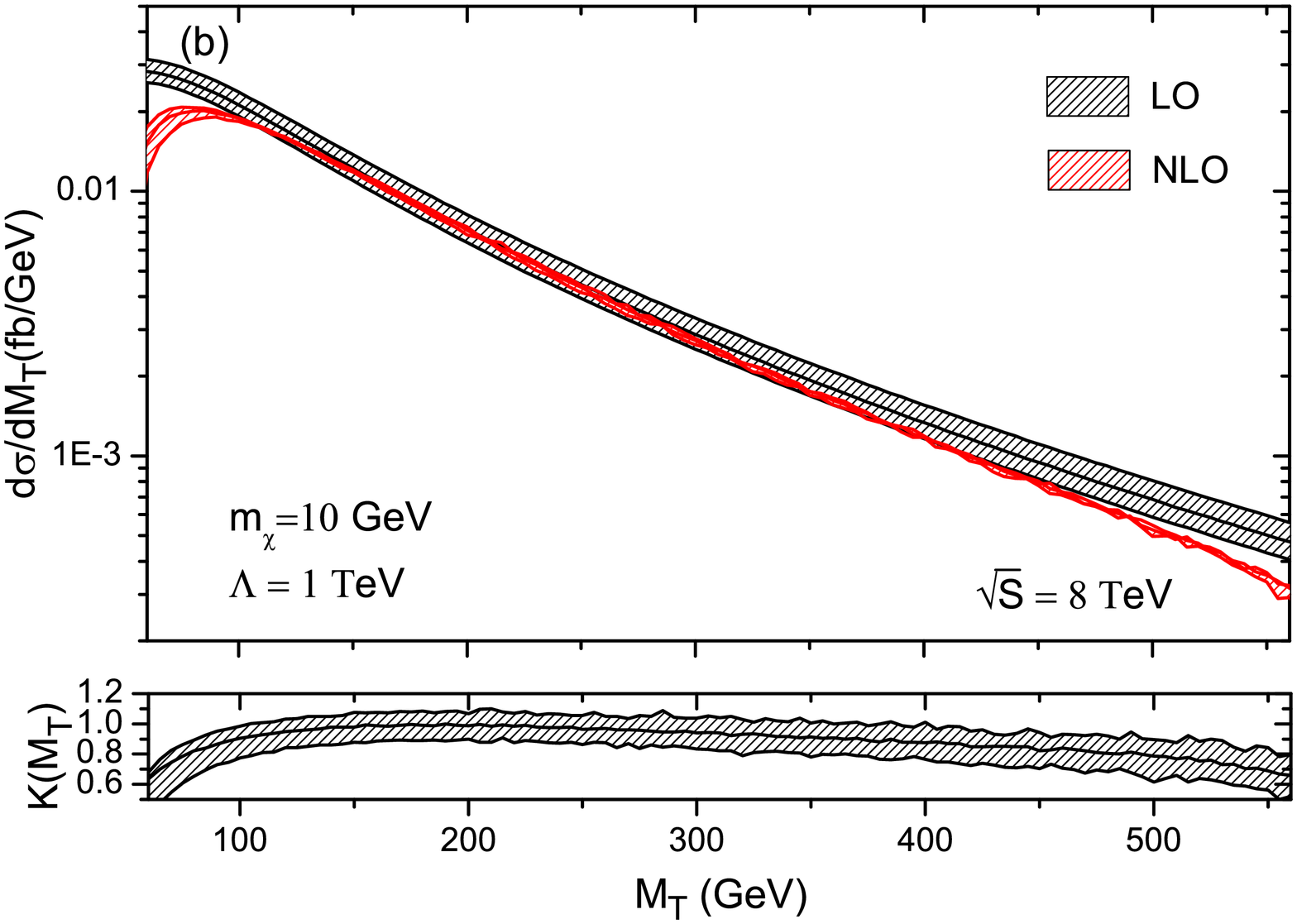}
\includegraphics[width=7.5cm,height=6cm]{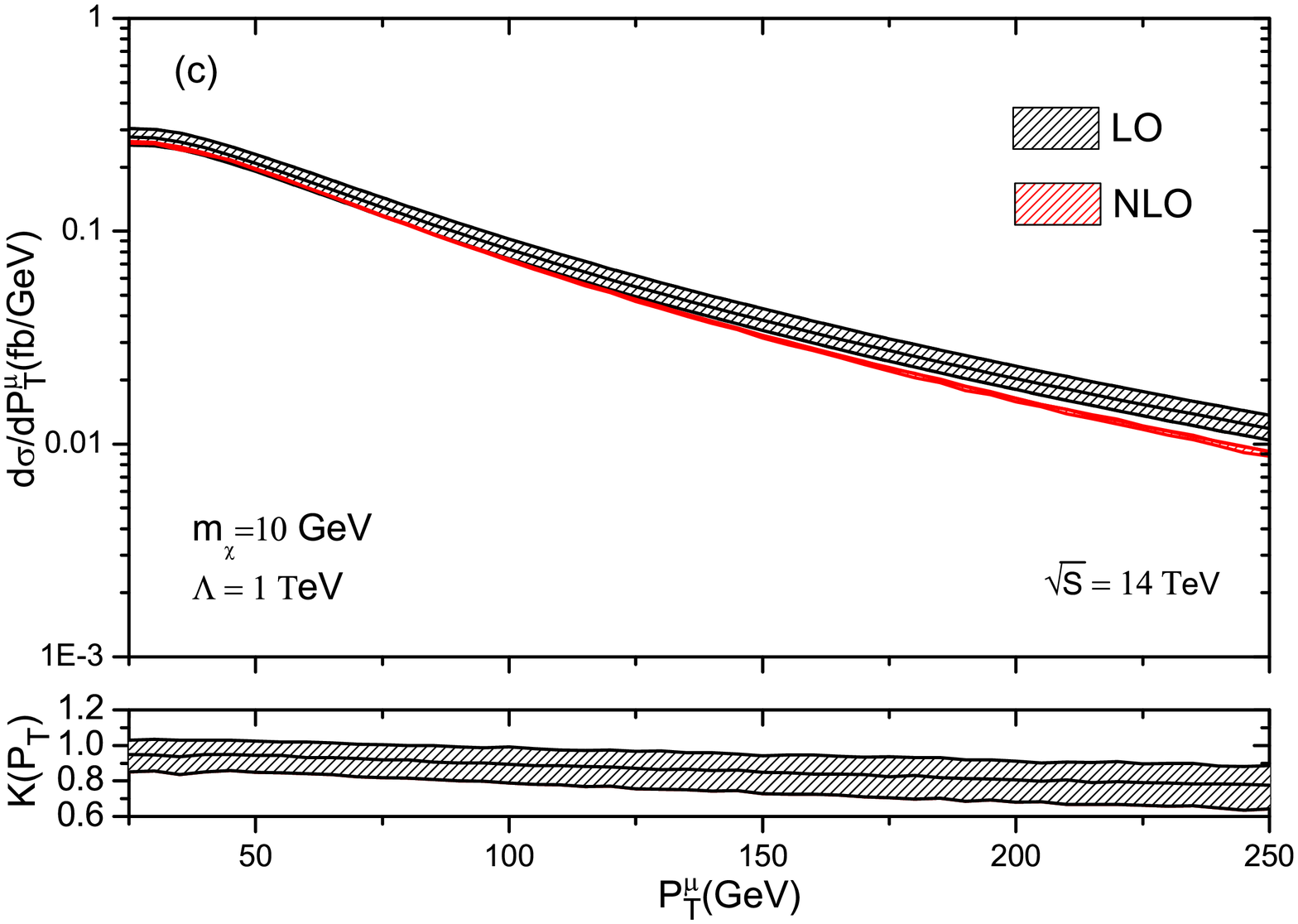}
\includegraphics[width=7.5cm,height=6cm]{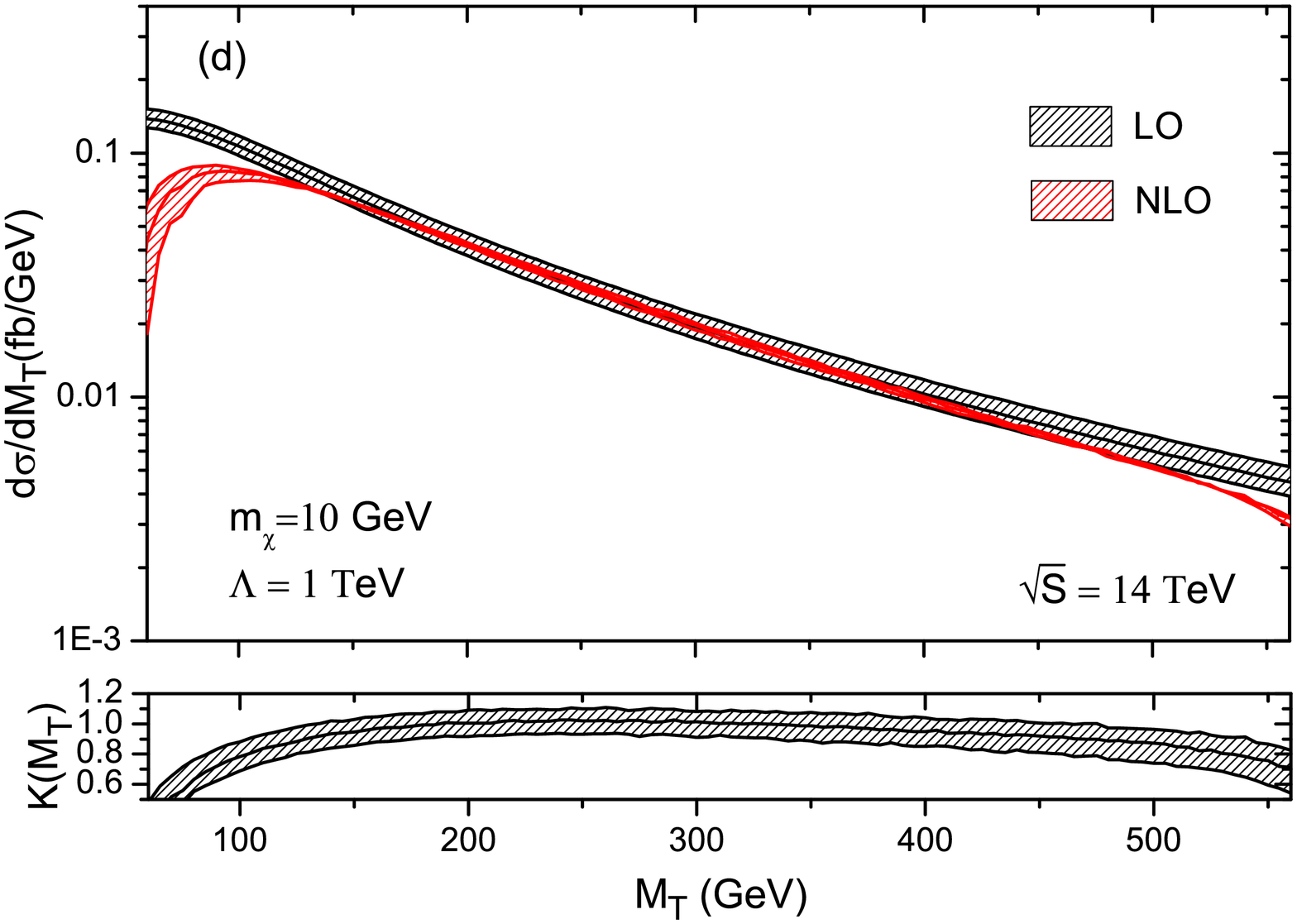}
\caption{\label{fig4} The LO and QCD NLO corrected distributions of
the transverse momenta of the final muon, transverse mass
$M_T$ and the corresponding $K$-factors for the $pp \to
\bar{\chi}\chi W^\pm + X \to
\bar{\chi}\chi\mu^\pm\stackrel{(-)}{\nu_{\mu}} +X$ processes at the
$\sqrt{s}=8~TeV$ and $\sqrt{s}=14~TeV$ LHC. The solid line indicates the differential
cross section obtained with the default scale $\mu=\mu_0$, the shaded band
represents the deviation from this scale when the scales are varied
by a factor of two in each direction. }
\end{figure}

In measuring the process $pp \to \bar{\chi}\chi+W^{\pm}+X$, the
finial $\bar{\chi}$ and $\chi$ particles are undetected as the
missing energy which will escape the detector without being
detected, and the $W$ boson is unstable and will decay to lepton and
neutrino. The neutrino is also not detected directly, and gives rise
to experimentally missing energy. We investigate the
kinematic distributions of final products after the subsequent
decays of $W$ gauge boson (i.e., $W^\pm \to
\mu^\pm\stackrel{(-)}\nu_{\mu}$). The SM leptonic decay branch
ratio of $W$ bosons is employed in further numerical calculations,
i.e., $Br(W^- \to \mu^-\bar{\nu}_\mu)= 10.57\%$ \cite{pdg}. The signature of
the $\bar{\chi}\chi W^\pm$ production at the LHC including their
subsequent decay can be written as
\begin{eqnarray}\label{channel}
pp \to \bar{\chi}\chi W^\pm + X \to
\bar{\chi}\chi\mu^\pm\stackrel{(-)}{\nu_{\mu}} +X.
\end{eqnarray}
Then that signal event is detected at the LHC as the detection of
one charged lepton $\mu^\pm$ plus missing energy.
In analogous to the definition in
CMS data analysis \cite{CMS:2013iea}, we select the events including
a muon with $p^{\mu}_T>25~GeV$. The parameter describing DM effect
is adopted in terms of the invariant transverse mass, which is
defined as
\begin{equation}\label{cuts}
M_T = \sqrt{2 p^\mu_T p^{miss}_T\left(1-\cos\Delta\phi_{\mu
D}\right)},
\end{equation}
where $p_T^{miss}=E_T^{miss}$ is the missing transverse momentum due to the undetected final
DM particle and neutrino, and $\Delta\phi_{\mu D}$ is the
azimuthal opening angle between the muon transverse momentum
direction and $\overrightarrow{p}_T^{miss}$. Another observable is
the transverse momentum of final muon. In
Figs.\ref{fig4}(a,b,c,d), we provide the LO and QCD NLO corrected
distributions of transverse momentum of muon from $W$ decay
($p_T^\mu$), invariant transverse mass $M_T$ and the corresponding $K$-factors
at the $8~TeV$ and $14~TeV$ LHC, respectively, where we take the
mass of DM and coefficients $\Lambda$ as $m_\chi=10~GeV$ and
$\Lambda=1~TeV$. Due to all the LO and NLO and $K$-factor
cureves for D8 coupling are nearly overlapped with the corresponding ones for D5 coupling,
we will don't distinguish between them in Figs.\ref{fig4}(a,b,c,d).
From these figures, we can see that the QCD
NLO correction obviously modifies the LO differential cross section.
For the transverse momentum distributions of muon, the $K$-factors
decrease slowly with the increment of $p_T^\mu$ from $25~GeV$ to
$250~GeV$ for both $\sqrt{s}=8~TeV$ and $\sqrt{s}=14~TeV$ LHC. For
the distributions of invariant transverse mass $M_T$, the QCD NLO corrected
differential cross section is very different from the LO
differential cross section at the lower and larger plotted $M_T$
regions. This is due to that the real gluon and real quark
radiations have an additional jet in the final states, which changes the
distributions of invariant transverse mass $M_T$.

\vskip 5mm
\section{Summary}
\par
The LHC provides an ideal facility to search for DM, offering
complementary results to those obtained from direct detection
experiments. Model independent searches for the pair production of
DM require some other visible activity in the event, e.g. jets,
photons, or vector bosons. In this paper we investigate the complete
QCD NLO corrections to the $\bar{\chi}\chi W^\pm$ associated
production at the LHC. We present the dependence of the LO and the
QCD NLO corrected cross sections on the factorization/renormalization
energy scale, and it shows that the dependence of the NLO cross
section on the factorization/renormalization scale ($\mu=\mu_f=\mu_r$)
are significantly reduced. We present the LO and the QCD NLO
corrected transverse momentum distributions of muon from
$W$ boson decays ($p_T^\mu$) and invariant transverse mass $M_T$.
We find that the QCD NLO radiative corrections obviously modify the
LO kinematic distributions, and the values of $K$-factor are obviously
related to the phase space and the kinematic observables. It shows that we should
consider the NLO QCD corrections in precision experimental data
analysis for this process.

\vskip 5mm
\par
\noindent{\large\bf Acknowledgments:} This work was supported in
part by the National Natural Science Foundation of China
(No.11205003, No.11305001, No.11275190, No.11375171, NO.11175001),
the Key Research Foundation of Education Ministry of Anhui Province
of China (No.KJ2012A021), the Youth Foundation of Anhui
Province(No.1308085QA07), and financed by the 211 Project of Anhui
University (No.02303319).

\end{document}